\documentclass{aastex62}
\usepackage{hyperref}
\usepackage{graphicx} 
\usepackage{listings}
\usepackage{amsmath}
\usepackage{booktabs}
\usepackage{threeparttable, tablefootnote}
\usepackage{algorithmic}
\usepackage[plain]{algorithm}
\usepackage{ulem}
\usepackage{appendix}

\submitjournal{PASP}

\newcommand{\BL}{{\textit{Breakthrough Listen }}}

\shorttitle{MeerKAT Target Selection}
\shortauthors{Czech et al.}

\newcommand{\UCB}{Department of Astronomy,  University of California Berkeley, Berkeley CA 94720}
\newcommand{\USQ}{Centre for Astrophysics, University of Southern Queensland, Toowoomba, QLD, Australia}
\newcommand{\UA}{University of Arizona: Steward Observatory, Tucson, AZ 85721, USA}
\newcommand{\pennst}{Department of Astronomy \& Astrophysics and Center for Exoplanets and Habitable Worlds
525 Davey Laboratory, The Pennsylvania State University, University Park, PA, 16802, USA}
\newcommand{\UT}{Dunlap Institute for Astronomy \& Astrophysics, University of Toronto, 50 St. George Street, Toronto, ON M5S 3H4, Canada}
\newcommand{\SWIN}{Centre for Astrophysics \& Supercomputing, Swinburne University of Technology, Hawthorn, VIC 3122, Australia}
\newcommand{\NIJ}{Department of Astrophysics/IMAPP,Radboud University, Nijmegen, The Netherlands}
\newcommand{\SETI}{SETI Institute, Mountain View, California}
\newcommand{\KZA}{University of Malta, Institute of Space Sciences and Astronomy}
\newcommand{\PWJD}{The Breakthrough Initiatives, NASA Research Park, Bld. 18, Moffett Field, CA, 94035, USA}
 
\newcommand{\SARAO}{South African Radio Astronomy Observatory, 2 Fir Street, Black River Park, Observatory, South Africa}

\begin{document}

\title{The Breakthrough Listen Search for Intelligent Life: MeerKAT Target Selection}


\author[0000-0002-8071-6011]{Daniel Czech}
\affiliation{\UCB}
\email{Corresponding Author: danielc@berkeley.edu}

\author[0000-0002-0531-1073]{Howard Isaacson}
\affiliation{\UCB}
\affiliation{\USQ}

\author{Logan Pearce}
\affiliation{\UA}

\author{Tyler Cox}
\affiliation{\UCB}

\author{Sofia Sheikh}
\affiliation{\pennst}

\author[0000-0002-7461-107X]{Bryan Brzycki}
\affiliation{\UCB}

\author[0000-0002-1691-0215]{Sarah Buchner}
\affiliation{\SARAO}

\author[0000-0003-4823-129X]{Steve Croft}
\affiliation{\UCB}

\author[0000-0003-3197-2294]{David DeBoer}
\affiliation{\UCB}

\author{Julia DeMarines}
\affiliation{\UCB}

\author{Jamie Drew}
\affiliation{\PWJD}

\author[0000-0002-8604-106X]{Vishal Gajjar}
\affiliation{\UCB}

\author{Brian Lacki}
\affiliation{\UCB}

\author{Matt Lebofsky}
\affiliation{\UCB}

\author{David H.\ E.\ MacMahon}
\affiliation{\UCB}

\author{Cherry Ng}
\affiliation{\UCB}
\affiliation{\UT}

\author{Imke de Pater}
\affiliation{\UCB}

\author[0000-0003-2783-1608]{Danny C.\ Price}
\affiliation{\UCB}
\affiliation{\SWIN}

\author[0000-0003-2828-7720]{Andrew P. V. Siemion}
\affiliation{\UCB}
\affiliation{\NIJ}
\affiliation{\SETI}
\affiliation{\KZA}

\author[0000-0001-8940-4228]{Ruby Van Rooyen}
\affiliation{SARAO}

\author{S. Pete Worden}
\affiliation{\PWJD}

\begin{abstract}
New radio telescope arrays offer unique opportunities for large-scale commensal SETI surveys. Ethernet-based architectures are allowing multiple users to access telescope data simultaneously by means of multicast Ethernet subscriptions. Breakthrough Listen will take advantage of this by conducting a commensal SETI survey on the MeerKAT radio telescope in South Africa. By subscribing to raw voltage data streams, Breakthrough Listen will be able to beamform commensally anywhere within the field of view during primary science observations. The survey will be conducted with unprecedented speed by forming and processing 64 coherent beams simultaneously, allowing the observation of several million objects within a few years. Both coherent and incoherent observing modes are planned. We present the list of desired sources for observation and explain how these sources were selected from the Gaia DR2 catalog. Given observations planned by MeerKAT's primary telescope users, we discuss their effects on the commensal survey and propose a commensal observing strategy in response. Finally, we outline our proposed approach towards observing one million nearby stars and analyse expected observing progress in the coming years.         
\end{abstract}

                    
\section{Introduction}
\label{sec:intro}

The objective of Breakthrough Listen's commensal SETI survey on the MeerKAT radio telescope array \citep{jonas2018, Booth2012} is to search for technosignatures in observations of one million nearby stars. This expansion of Breakthrough capabilities complements the initial survey described by \cite{Isaacson2017} that established the list of 1702 nearby stars, 100 galaxies and a survey of the galactic plane of the Milky Way. While the primary survey has been documented in  \cite{Enriquez2017} and \cite{Price2020}, observations and analyses of targets of opportunity have complemented the original survey design \citep{Gajjar2018, Enriquez2019, Price2019_frb}, and smaller-scale observing projects have added to the original target list \citep{Sheikh2019}. 

Customized hardware at the Green Bank Telescope (GBT) \citep{MacMahon2018} and Parkes Observatory  \citep{Price2018_parkes_instr} has enabled searches for narrow band radio emission via broadband voltage capture and subsequent data reduction, resulting in high resolution data products that are suitable for narrowband signal searches \citep{Lebofsky2019}. The primary star survey was revisited by \cite{Wlodarczyk-Sroka2020} showing that the first few analyses of Breakthrough Listen (BL) radio data have surveyed many background stars in addition to the targeted nearby stars. An optical survey of the nearby star sample with the Automated Planet Finder at Lick Observatory gives BL a reach beyond the radio enabling SETI searches in high resolution optical spectroscopy \citep{Isaacson2019,Lipman2019}. As the fields of astronomy and machine learning become more intertwined, enhanced searches for narrowband radio signals \citep{Brzycki2020} will add to those utilizing machine learning to search for fast radio bursts \citep{Zhang2018}.  

The work of BL is built upon decades of previous SETI programs on radio telescopes. A commensal SETI search was proposed for the Allen Telescope Array (ATA) as early as 2006, featuring the capability to beamform concurrently on 16 targets \citep{DeBoer2006}. The science goals for the ATA originally included the observation of $10^6$ stars with sufficient sensitivity to detect the Arecibo planetary radar out to 300\,pc \citep{Gutierrez2010}. SETI searches of known exoplanet systems have been conducted with the ATA \citep{HarpG2016}, while the Very Large Array (VLA) observed the galaxies M\,31 and M\,33 \citep{Gray2017}. The Murchison Widefield Array (MWA), observing in the frequency range $70 - 300$\,MHz, has performed serendipitous SETI observations (for example of the interstellar object 'Oumuamua  \citep{Tingay2018}, using data acquired for other observations). Plans are also underway for a pilot program to perform beamformed SETI searches at the MWA \citep{beardsley2020}. A commensal SETI survey with the VLA is in development, using an Ethernet-based commensal observing system \citep{Hickish2019}. 

A number of radio frequency SETI experiments have also been conducted with single-dish telescopes in recent years. \cite{Pinchuk2019} observed twelve exoplanetary systems, including TRAPPIST-1, for a total of two hours with the GBT at L-band. The FAST radio telescope in Guizhou, China has conducted its first SETI observations using a 19-beam receiver from $1 - 1.5$\,GHz \citep{Zhang2020} and will continue to conduct commensal SETI experiments. These studies build on the legacy of other single dish SETI experiments including  \cite{Tarter2001,Werthimer2001,Tarter2011,Korpela2011} and \cite{Siemion2013}. 

Optical and infrared technosignature searches have grown in number in recent years. \cite{Maire2019} conducted a search for nanosecond-scale transients, observing 1280 objects from $950 - 1650$\,nm with sensitivity to optical pulses shorter than 50\,ns. Boyajian's Star (KIC\,8462852), with its anomalous optical dimmings, has been the target of optical SETI searches \citep{Lipman2019} and has been observed in the radio spectrum by the ATA \citep{Schuetz2016}. Infrared searches have been conducted using WISE data to investigate the existence of extraterrestrial civilizations with large energy supplies \citep{Griffith2015}. The Pulsed All-sky Near-infrared Optical SETI (PANOSETI) observatory is a future instrument which will search for transient technosignatures and astrophysical phenomena \citep{WrightS2019}. Its design enables simultaneous full sky coverage and the capability to detect pulsed optical signals in the nanosecond to second range.  The VERITAS project \citep{Holder2006} is currently collaborating with BL on a survey of nearby stars searching for coincident pulsed signals of nanosecond duration amidst its study of high-energy astrophysics via gamma rays, using four 12-m optical reflectors at the Fred Lawrence Whipple Observatory.

As observational SETI searches have advanced with improvements in hardware and data capacity, theoretical studies of occurrence and probablity of detecting intelligent life beyond Earth have also proliferated \citep{Gray2020,Grimaldi2018,Grimaldi2018b}. However, we maintain that the only real measure of intelligent life beyond Earth is an observational one. 

The principal aim of this work is to compile a target list of nearby stars that BL will aim to observe during commensal observations with the MeerKAT radio telescope. MeerKAT is located in a remote radio-quiet region of South Africa, and consists of 64 offset-Gregorian antennas \citep{jonas2018}. UHF and L-band receivers are available, with S-band receivers undergoing commissioning. The different components of the telescope are connected via an Ethernet network, allowing multiple users of the telescope to receive data simultaneously by subscribing to multicast data streams. Along with its wide primary field of view and radio-quiet southern hemisphere location, this capability renders MeerKAT well-suited for a commensal SETI survey.

Using Gaia DR2 \citep{gaia2018} we generate our primary list of approximately 26 million stars, providing ample targets to observe within the confines of commensal observing. The outline of the paper is as follows: Section \ref{methods} discusses how we selected targets from Gaia DR2 and describes their demographics. Our observing strategy given limitations imposed by the Large Science Projects (LSPs hereafter) is discussed in Section \ref{discussion}. Future observing progress is estimated and discussed in Section~\ref{optimal_timeline} and the design of the real-time target selection software is discussed in Section~\ref{target_selection}.  In Section~\ref{figures_of_merit} we compare the proposed commensal survey on MeerKAT with prior SETI surveys.  

\section{Methods}
\label{methods}
\subsection{Database Sources}

   We use the Gaia DR2 stellar database \citep{gaia2018} in order to produce the 26 million star catalog, which will comprise the full primary sample of the Breakthrough Listen-MeerKAT (BL-MeerKAT) survey. We aim to observe at least 1M stars with MeerKAT, more stars by a factor of 1000 than the next largest survey \citep{Isaacson2017}. We present 26M stars in our catalog to ensure that we have a sufficient number of stars on which to form 64 simultaneous beams on any part of the sky.
   
\begin{figure}[htp]
    \centering
    \includegraphics{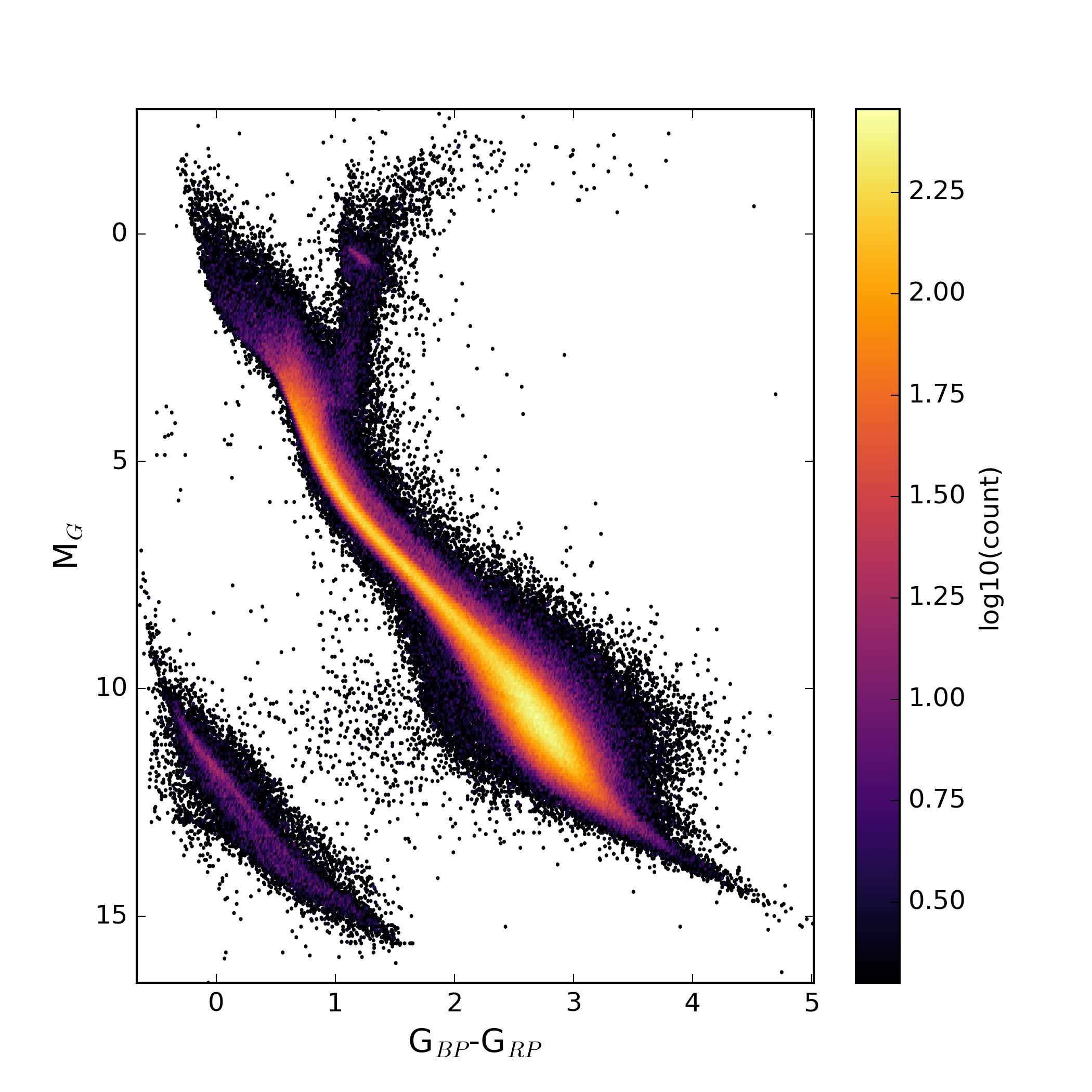}
    \caption{Color magnitude diagram of one million nearby stars. The quality cuts on the Gaia DR2 data concentrate the sample on parts of the diagram known to be populated and to be robust distance measurements. Bright colors represent a high density of stars, dark colors, low density. }
    \label{fig:CMD_1M}
\end{figure}

   Gaia DR2 allows for uniform sampling over the entire celestial sphere, with well documented errors on stellar brightness, systemic radial velocity and parallax \citep{Luri2018}, from which distance can be calculated \citep{Bailer-Jones2015}. We use quality control cuts from \cite{gaia2018_HR_diagrams} on parallax, flux, and astrometric excess in order to produce a well understood catalog of stars with small errors in distance as calculated from parallax. The astrometric excess is related to the quality of the astrometric solution, with larger values meaning lower quality. While creating a target list with high integrity, brighter stars will be selected more commonly than fainter stars and multi-star systems, which have large parallax errors and will therefore tend to be dismissed \citep{Bailer-Jones2018} hereafter BJ18. This decision leads to a catalog with well understood limitations that retains stars which fall within the parameter space of color-magnitude diagrams that is known to be populated (Figure \ref{fig:CMD_1M}).
   
\begin{figure}[htp]
    \centering
    \includegraphics[width=0.75\textwidth]{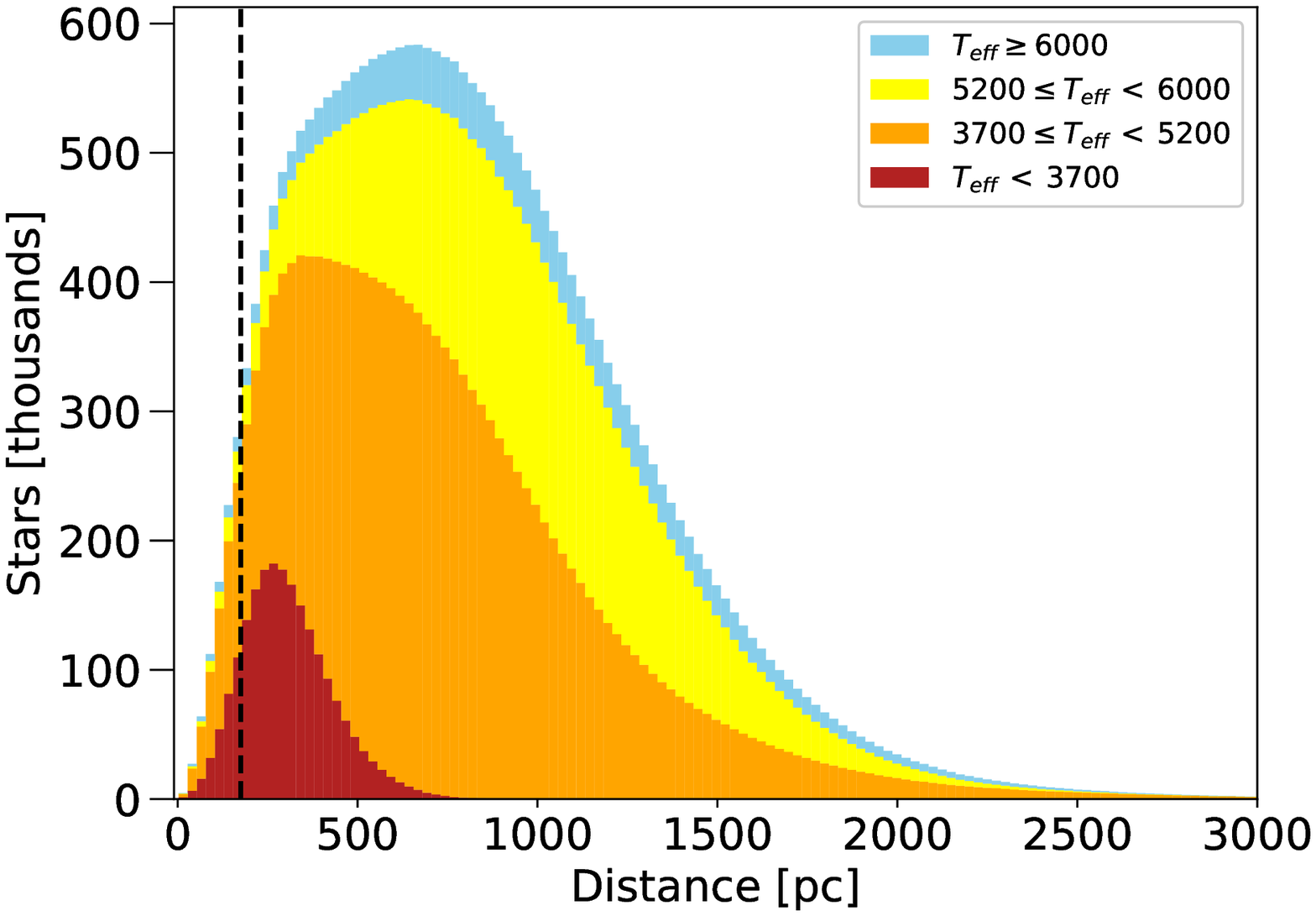}
    \caption{Distances to our star sample as calculated from Gaia DR2 parallax measurements. The one million nearby stars, which are of primary interst to us, appear to the left of the vertical dotted line. The most distant of the nearest one million stars is 175.447\,pc away. The median of the distribution is at approximately 771\,pc, and 30,200 stars are at distances greater than 3000\,pc. }
    \label{fig:1M_dist_teff}
\end{figure}

   The complete list of Gaia quality metrics including cuts on fractional errors in parallax, flux values in the Gaia bandpasses ($G,~ G_{RP},~G_{BP}$), color excess and astrometric fit quality is provided in Table \ref{table:gaia_quality_cuts}. The quality cuts on parallax over parallax error ensure all of our calculated distances have high integrity. The three cuts on magnitude are used to remove variable stars. Cuts on color excess are made to remove stars that are blended with nearby stars, resulting in poor parallax and distance measurements due to contamination. The number of visibility periods required for a five parameter astrometric solution is 6, whereas we choose a more conservative 8 periods because distance integrity is important to us.  The limit on astrometric chi-squared will remove binary systems from our sample. Currently, binary stars have poor distance estimates. Collectively, these metrics filter out stars that fall into non-physical areas of the HR diagram \citep{gaia2018_HR_diagrams} and ensure high quality distance measurements. The chosen metrics result in a stellar sample of about 32 million stars (32167216) over the entire celestial sphere, of which approximately 26 million (26432021) are visible to MeerKAT. As a result of our stricter quality constraints (described in Table~\ref{table:gaia_quality_cuts}), our sample contains fewer stars than the catalogue compiled by BJ18. For example, approximately 25\% of the stars from the BJ18 catalogue, within 500\,pc, appear in our sample. In order to simulate our observing progress, we collect anticipated sky pointings representative of those that will be conducted by MeerKAT's primary observers (see Table \ref{table:lsps}) and determine the stars that fall within MeerKAT's primary field of view during the first 6 months of observations. Ninety-five percent of the stars in the volume-complete sample (containing Gaia objects that meet our quality cuts on distance and magnitude) are within 2000\,pc, with a median distance of approximately 771\,pc (Figure \ref{fig:1M_dist_teff}). Additional details on the steps from target to selection to beamforming will be provided in Section \ref{strategies}. The full catalog is available for download at https://seti.berkeley.edu/meerkat\_db/BL\_MeerKAT\_target\_list\_2021.csv.gz.

\begin{table}[h]
    \centering
    \caption{List of quality metrics used to select the stars comprising the primary star sample. Full descriptions and SQL queries are available in Appendix B of \cite{gaia2018_HR_diagrams} }
    \begin{tabular}{|l|c|c|}
    \hline
    \hline
    Variables &  SQL Query column & Value \\
    \hline
  $f_{true}$ & parallax\_over\_error & $ > 20$ \\ 
  $\sigma_{G} $  & phot\_g\_mean\_flux\_over\_error & $ > 50$ \\  
  $\sigma_{G_{RP}} $ & phot\_rp\_mean\_flux\_over\_error & $> 20$ \\
  $\sigma_{G_{RP}} $ & phot\_bp\_mean\_flux\_over\_error & $> 20$ \\
  $(I_{BP} + I_{RP})/I_{G} $ & phot\_bp\_rp\_excess\_factor      & $ < 1.3 + 0.06 \times (G_{BP} - G_{RP})^{2}$ \\
  $(I_{BP} + I_{RP})/I_{G} $ & phot\_bp\_rp\_excess\_factor      & $ > 1.0 + 0.015 \times (G_{BP} - G_{RP})^{2}$ \\
  -- & visibility\_periods\_used        & $\geq 8$ \\
  $\chi^{2},~\nu$  & astrometric\_chi2\_al $/$ (astrometric\_n\_good\_obs\_a -5) & $ < 1.44 * max(1,e^{-0.4(G-19.5)}) $ \\

    \hline
    \end{tabular}
\label{table:gaia_quality_cuts}
\end{table}    

\subsection{Calculating Stellar Distance}

Our calculation of the distances to the stars in our sample follows the procedure outlined by BJ18, which uses Gaia parallax, parallax error, and galactic latitude and longitude to calculate likelihood functions for the distances to stars. Both BJ18, and \cite{Luri2018} promote the calculation of distance from Gaia parallax with a full Bayesian approach in order to avoid overestimating the distance, which  can occur when the simple inverse of parallax is used. The likelihood of the parallax measurement for a given distance is described by a Gaussian distribution, with some variance about the measurement. Fractional parallax errors are used as a metric for the quality of the parallax measurements, with small values indicating higher quality.  Priors are implemented such that for more precise parallax measurements, the prior is less influential, and when the parallax measurement has lower quality the prior is more meaningful in the distance calculation. BJ18, Figure 5, provides representative distance posteriors and priors.

\cite{Bailer-Jones2015} and BJ18 recommend using a Markov-Chain Monte Carlo (MCMC) algorithm to estimate the distance and corresponding errors. Testing  several statistics on the posteriors as distance estimators, BJ18 finds the mode of the distribution is the best distance estimator. The 95\% credible intervals are preferred over the standard deviations as uncertainties in distance, since a standard deviation has the potential to allow non-physical negative distances. Because many of the posteriors are not well approximated as Gaussian, using a simple  $\sigma= FWHM/2.355$ is not accurate in most cases.  We create a custom MCMC sampler in Python, based on the R tutorial used in \cite{Bailer-Jones2015}. The final jump rate found an acceptance rate of 40\% and 5000 samples were used balancing the robustness of our results with run time.  As the fractional error decreases, the resulting parallax posterior approaches the shape of the prior and no longer reflects the data. This results in difficulty in defining the uncertainty. Our data quality cuts ensure that the distance uncertainty is robust. In comparison with the distances obtained by BJ18 for stars within 500\,pc, our estimates differ by a median of +0.98\% and a maximum of +1.69\%.

Spectral type is not directly considered in this selection process, but since Gaia is a magnitude-limited survey, some nearby faint, cool stars will be missed and some brighter, hotter, more distant stars will be observed instead. Early Gaia Data Release 3 (EDR3) is providing improved distance values for 1.47 billion stars \citep{bailer-jones2020}, but we anticipate this having a small effect on our sample. The EDR3 analysis of stars within 100 pc, which is 92\% complete down to spectral type M9 \citep{gaia_gnsc_2020}, contains  331,312 stars. Based on our DR2 sample, which contains 224,401 stars within 100 pc, but includes strict data quality cuts, we anticipate fewer than the 106,911 star difference will be added to our sample.  Figure \ref{fig:1M_dist_teff} shows the distribution of distances and stellar effective temperature from Gaia DR2.

\subsection{Supplemental Catalogs}
\label{special_targets}

In addition to the star sample presented in Section~\ref{methods}, it is expected that observations of objects in additional and supplemental target lists will be desired. Here we consider several examples.  

\subsubsection{Earth Transit Zone Observations}
The Earth Transit Zone (ETZ) is a region bracketing the ecliptic from which extraterrestrial observers would be able to detect transits of the Earth in front of the Sun, and perhaps even characterize the biosignatures in our atmosphere \citep{Kaltenegger2020}. Non-ETZ extraterrestrial observers would not have access to such information. For decades, humans have preferentially directed strong, clearly-artificial radar transmissions in our ecliptic to study the Solar System. These transmissions would only be detectable by extraterrestrial observers located in the ETZ. 
While the ETZ has been theorized as a promising region for SETI searches for decades \citep[e.g.,][]{filippova1988ecliptic, castellano2004visibility, shostak_villard_2004}, methodical targeted searches of this region have only been undertaken recently. \citet{Sheikh2020} observed 20 targets in the ETZ, but the MeerKAT program could greatly expand this number. MeerKAT has the capability to observe up to 186,780 Gaia DR2 objects in the ETZ within 600\,pc. These targets were chosen to have astrometric excess noise values of less than 5$\sigma$ to guarantee real sources. Subjecting this sample to the same quality metrics as our primary sample, we see that 18,656 stars from the ETZ are already included in our full sample.

\subsubsection{Exotica}

Occasionally there arise targets for which observations are urgently desired. For example, when the interstellar object 'Oumuamua passed through the solar system, BL conducted dedicated observations of it using the GBT \citep{enr2018}. Another example is BL's campaign to observe Boyajian's Star for optical laser line emission, using the Automated Planet Finder at the Lick Observatory \citep{Lipman2019}.

As the survey on MeerKAT is completely commensal in nature, immediate follow-up observations of newly discovered exotica are not likely to be possible. Dedicated primary time on telescopes such as Parkes and GBT are a more practical alternative. Nevertheless, a mechanism to specify a target for immediate attention is beneficial. As discussed in Section~\ref{target_selection}, a special category of ad-hoc targets will be created for this purpose, and will be prioritised over all others when triaging targets during commensal observing.

Non-urgent exotica targets will be queued for observation along with our full  star catalog. \cite{Lacki2020} provides an extensive list of 737 exotica targets from which to choose. Each of these are sorted into one of four separate categories (Prototype, Superlative, Anomaly and Control) which are discussed further in \cite{Lacki2020}. 

As there are relatively few of these objects, the likelihood of individual unique exotica objects falling within the primary field of view during commensal observations is low. Therefore, in future work, we will compile a catalog containing multiple examples of each particular category of exotica. For example, instead of a single object with the most superlative characteristics in a particular category, a ranked list of superlative objects will be provided. Targets will be drawn from several existing catalogs describing relatively exotic but still fairly common objects such as a catalog of cataclysmic variables \cite{downes2006} or a catalog of interesting radio pulsars \cite{manchester2005}. We will also develop the capability to beamform on asteroids and minor bodies that fall within the primary field of view during commensal observations.

\begin{figure}[htp]
    \centering
    \includegraphics[width=0.9\textwidth]{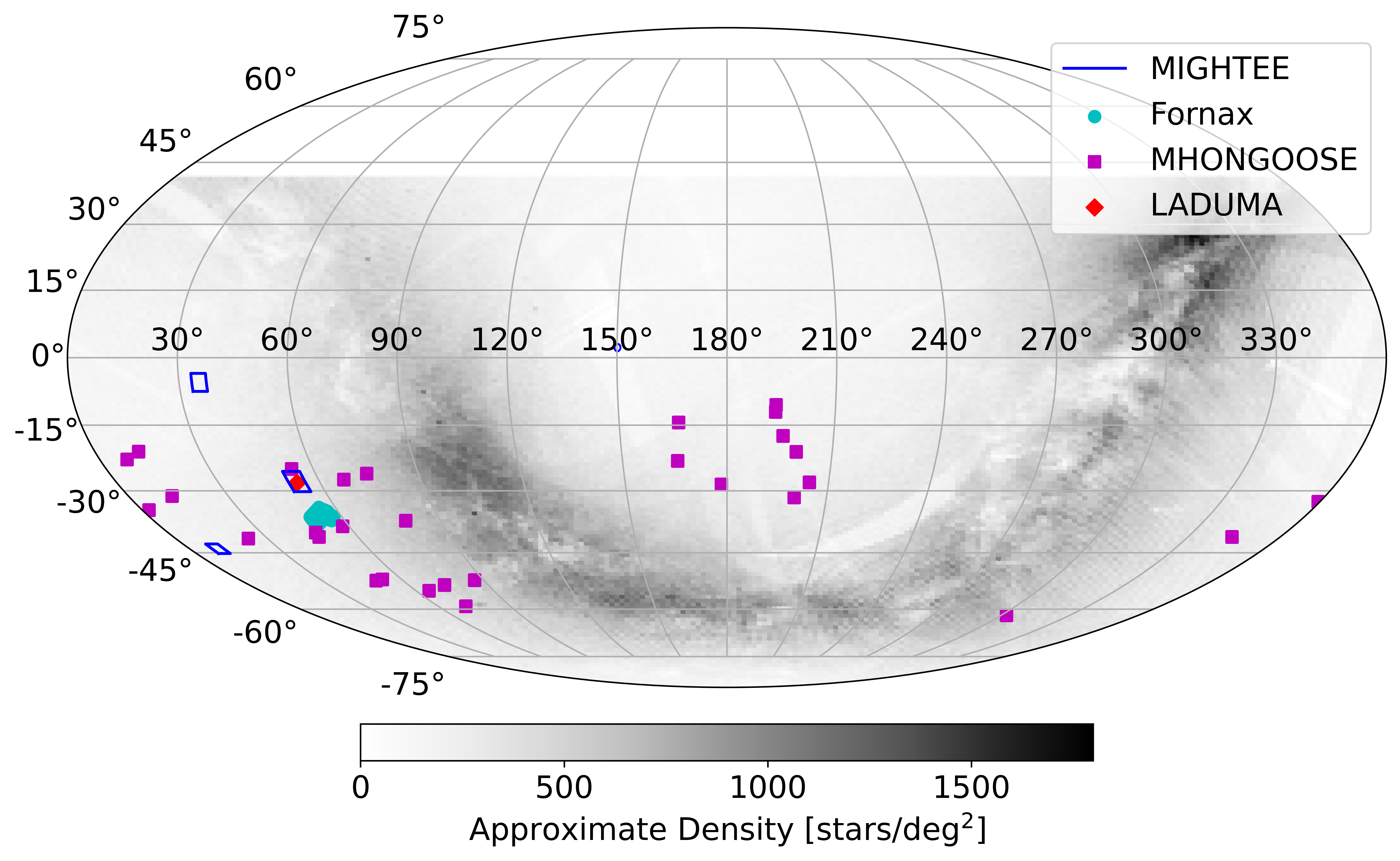}
    \caption{Sky positions of several of the major LSPs. The grey background shows the density of stars, as drawn from our full catalog.}
    \label{fig:lsp_pointings}
\end{figure}

\section{Discussion}
 \label{discussion}
 
\subsection{Observing Capabilities and Limitations of the BL System at MeerKAT}

Here we provide an abridged description of BL's observing system at MeerKAT and its capabilities, limitations and current status. Complete technical details of the system will be provided in an upcoming instrumentation article. 

Breakthrough Listen's system will consist of 128 processing nodes and 8 storage nodes (amounting to 1.584\,PB of usable storage). An online processing pipeline is in development. Data is ingested from the F-engines and buffered into NVMe modules in the RAW format, with a bit-depth of 8. The RAW format used is broadly similar to that described by \cite{Lebofsky2019}, albeit with modifications to handle data from multiple antennas. Further channelisation to a resolution of 1Hz will take place. Both an incoherent mode and a coherent beamforming mode will be implemented (see Sections~\ref{incoherent-beam} and \ref{beamforming} for further details).

The NVMe buffers are selected to have sufficient capacity to hold up to 5 minutes' worth of raw voltage data at the maximum data rate we expect to receive from the MeerKAT F-engines. The 5-minute duration is chosen for consistency with prior Breakthrough Listen observing campaigns, in which the on-off observing strategies make use of individual 5-minute pointings. For example, \cite{Price2020}, \cite{Perez2020}, \cite{Brzycki2019} and  \cite{Enriquez2017} use individual pointing durations of 5 minutes. Selecting buffers of sufficient size to store raw voltages for a full 5 minute pointing allows us to draw comparisons with our prior work more easily. In addition, it makes it easier to adapt our existing software tools for single-dish radio telescopes to array radio telescopes like MeerKAT.

The incoherent and coherent beamforming stages will be followed by SETI search components, detecting narrowband drifting signals using TurboSETI \citep{enr2019a} for example. The volume of data recorded by the system is too large to store in its entirety. Instead, detected signals of interest will be excised from the raw data (in the form of time-frequency ``postage stamp" regions) and saved to the storage nodes. The rest of the raw data will be discarded. Budgeting for an average frequency by time range of 6\,kHz by 300\,s, each saved region would amount to 450\,MB. Over 3 years, we expect to be able to save approximately 3215 of these regions per day on average, assuming 1.584\,PB of storage and that each region is never deleted. 

\begin{figure}[htp]
    \centering
    \includegraphics[width=0.8\textwidth]{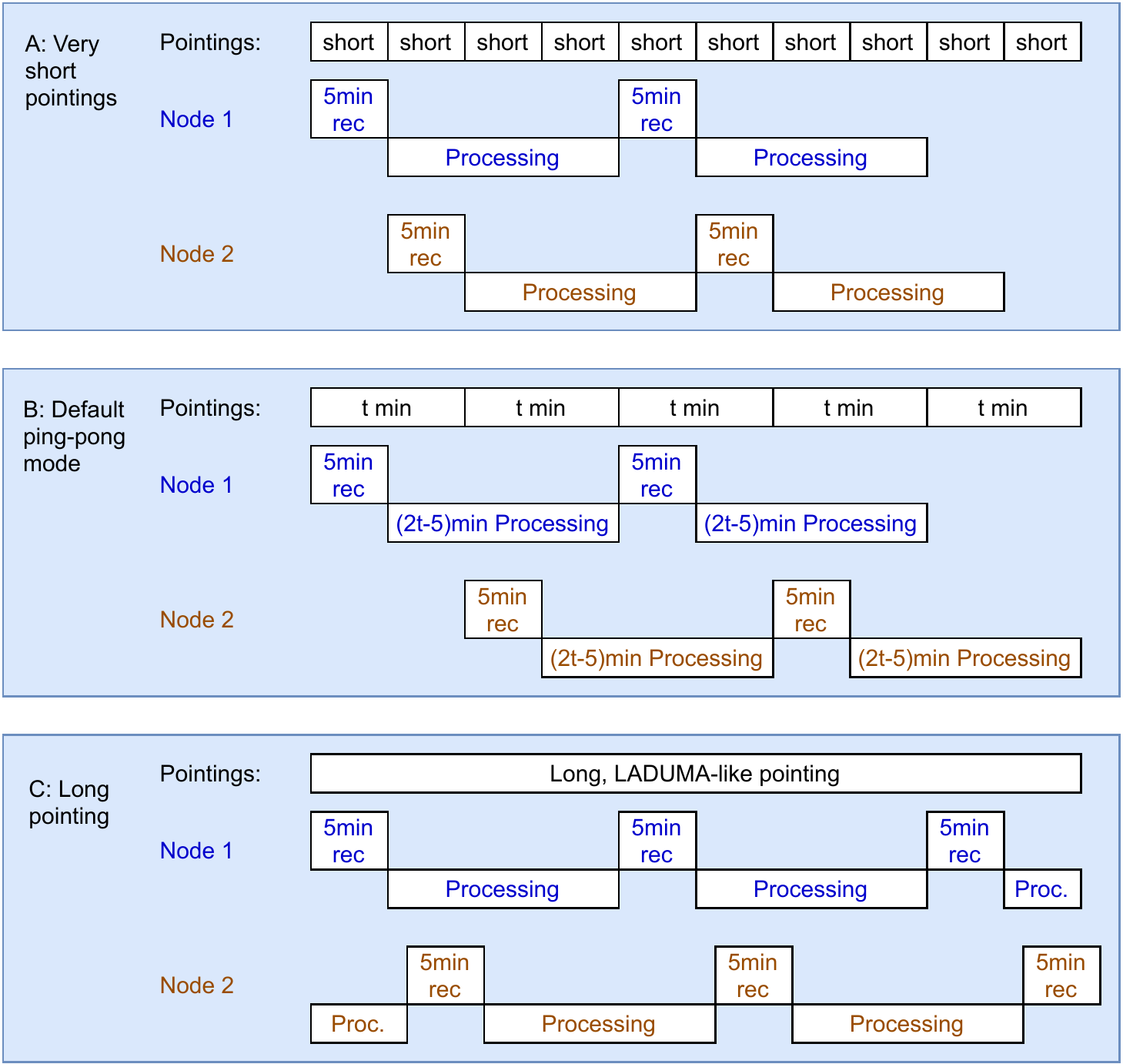}
    \caption{Three scenarios that could be encountered during commensal observations. In scenario A, the primary observer conducts a series of pointings shorter than the total processing time required. Under this scenario, some pointings will have to be ignored. Thankfully, such short pointings ($\leq 5$ minutes) appear to be rare. By far the more common scenarios are those depicted in the middle and bottom of the figure. In scenario B, the default ``ping-pong" mode, the pointings (of duration $\tau_{pointing}$) are longer and allow $2\tau_{pointing} - 5$ minutes of processing time when both banks are operating. By alternating between banks of processing nodes, data from each pointing can be recorded and processed. In the final scenario, illustrated at the bottom of the figure, $\tau_{pointing} \gg \tau_{processing}$, allowing a more flexible recording and processing approach for both banks.}
    \label{fig:scheduling}
\end{figure}

The system at MeerKAT will be divided into two banks of 64 processing nodes. Each bank will be capable of recording the full bandwidth produced by the F-engines, from all antennas (in all observing bands). One bank will record and begin processing one buffer's worth of data while the other completes processing the previous buffer's worth of data, in a so-called ``ping-pong" fashion. The processing pipeline, and in particular, the SETI search components, will be reconfigured repeatedly over the lifetime of the system. Therefore, the processing time required is likely to vary. In addition, the time budget available for processing is dependent on the duration of each pointing conducted by the primary observer. Examples of each scenario are provided in Fig.~\ref{fig:scheduling}. 

The BL system at MeerKAT is being developed in a phased approach. At the moment, 50\% of the computing and storage capacity has been installed at the Karoo Data Rack Area on site at the telescope. The remaining 50\% will be installed in mid-2021. Work is ongoing to develop and commission the processing pipeline using the existing hardware already installed on site. 
Currently, we are able to receive and buffer raw voltage data from the MeerKAT F-engines at the maximum expected datarate. Data is recorded during commensal observations in a fully automated, unattended manner. Due to its relative simplicity, the incoherent sum mode will be commissioned first, followed by the coherent beamforming mode. We anticipate that both the incoherent and coherent beamforming modes will be operational by mid-2021, when the remaining 50\% of the hardware will be installed. 

\subsection{Observing Strategy}
\label{strategies}

 The Large Survey Projects listed in Table~\ref{table:lsps} will dictate the pointings for the majority of MeerKAT's observing time, and special consideration must be given to the impact of each LSP's observing strategy on our survey.  Each LSP has unique science goals and their planned observations are optimised to fulfill those objectives. At one extreme, lies the LADUMA LSP \citep{bly2016} which will spend 3424\,h observing a single pointing of one square degree, while at the other, the LSPs MeerTIME \citep{bai2016} and TRAPUM \citep{stappersb2016} will observe thousands of unique pointings, some repeatedly, to achieve their pulsar science goals. As a result, our proposed SETI survey's commensal observing strategy will need to be adapted accordingly.  The LSPs with the largest sky coverage that observe the accessible sky most uniformly will be most beneficial for the BL-MeerKAT survey to approach its goal of observing one million nearby stars. 

 When observing commensally with LSPs undertaking long dwell times in narrow regions of the sky, we will probe the nearest stars in those fields with increased sensitivity. Details of observation strategy including coherent and incoherent beamforming are discussed in Sections \ref{beamforming} and \ref{incoherent-beam}. Details of the current LSPs including chosen receivers and time on sky are collected in Table \ref{table:lsps}, and sky coverage for several of them is illustrated in Fig.~\ref{fig:lsp_pointings}. 

{\renewcommand{\arraystretch}{1.5}
\begin{table}[h!]
\caption{MeerKAT Primary Observers}
\begin{threeparttable}
\begin{tabular}{p{0.23\textwidth}p{0.1\textwidth}p{0.52\textwidth}} 
    \toprule
    \textbf{Project} & \textbf{Band} & \textbf{Description} \\
    \midrule

    LADUMA \newline \citep{bly2016} & UHF, L & LADUMA (Looking at the Distant Universe with the MeerKAT Array) is a deep neutral hydrogen survey to detect emissions of redshifts up to $z=1.4$. Up to 3424 hours will be spent observing a single pointing (03:32:30.4, -28:07:57), divided between L-band (up to 333 hours) and UHF-band (up to 3091 hours). \\
    
    MeerTIME \newline \citep{bai2016, bailes2020} & UHF, L, S\tnote{1} & MeerTIME will conduct extensive pulsar observations including regular timing of pulsars. Up to a potential 5400 hours will be divided among pulsar binaries, a pulsar timing array, globular clusters and MSPs. In addition, pulsar glitching and radio magnetars will be monitored. MeerTIME will also work together with TRAPUM and observe pulsars the latter discovers. \\ 

    TRAPUM \newline \citep{stappersb2016} & L, S\tnote{1}  & TRAPUM\tnote{6}~ (Transients and Pulsars with MeerKAT) will conduct searches for pulsars and transients with durations ranging from microseconds to seconds. Up to 976 hours will be used for observations of Fermi sources, nearby galaxies, globular clusters, and SNR, PWNe, TeV and $\gamma$-ray sources. \\
    
    Fornax \newline \citep{ser2017} & L & The MeerKAT Fornax survey\tnote{2}~ consists of a mosaic of 91 pointings covering the Fornax galaxy cluster, observing for up to 900 hours in total.\\
    
    MIGHTEE \newline \citep{jarvis2017} & L, S\tnote{1} & MIGHTEE (MeerKAT International Gigahertz Tuned Extragalactic Exploration) is a radio continuum survey that will observe four extragalactic deep fields (amounting to 20 square degrees) for up to 1920 hours in total (including overhead). \\
    
    ThunderKAT \newline \citep{fender2017} & L (and others) & ThunderKAT (The Hunt for Dynamic and Explosive Radio Transients with MeerKAT) will conduct both primary and commensal observations to search for explosive radio transients (including X-ray binaries, cataclysmic variables, short gamma-ray bursts and type Ia supernovae as stated on the website\tnote{4}~).\\

    MALS \newline \citep{Gupta2017} & UHF, L & MALS\tnote{5}~ (the MeerKAT Absorption Line Survey) will search for HI and OH absorption lines (redshift $0<z<2$) for up to 1655 hours in total. As mentioned by \cite{Gupta2017},\newline
    \textbullet \hspace{2mm} 70\% of the L-band pointings will be in the range $-30\degr < \delta < +40\degr$ \newline
    \textbullet \hspace{2mm} 70\% of the UHF-band pointings will be in the range $-30\degr < \delta < +40\degr$ \newline \\

    MHONGOOSE \newline \citep{deblock2016} & L & MHONGOOSE\tnote{6}~ (MeerKAT Observations of Nearby Galactic Objects - Observing Southern Emitters) will observe the neutral hydrogen distribution of 30 nearby galaxies for up to 55 hours each. \\

    Max Planck S-band System \newline \citep{kramer2016} & L, S\tnote{1}~ & The MPIfR is providing an S-band receiving system which will enable a survey of the inner galaxy, the details of which are forthcoming. \cite{kramer2016} discuss a possible survey covering $|l| < 30\degr$ and $|b| < 0.5\degr$. \\

    Open Time & UHF, L, S\tnote{1} & 28\% of MeerKAT's observing time will be dedicated towards open time calls for proposals.\\

    Director's Discretionary Time & UHF, L, S\tnote{1} & 5\% of MeerKAT's observing time will be dedicated towards Director's Discretionary Time.\\

\bottomrule
\end{tabular}
\begin{tablenotes}\footnotesize
\item [1] S-band receivers are currently undergoing installation and commissioning. 
\item [2] http://www.trapum.org
\item [3] https://sites.google.com/inaf.it/meerkatfornaxsurvey/goals-design
\item [4] http://www.thunderkat.uct.ac.za/ 
\item [5] https://mals.iucaa.in/survey
\item [6] https://mhongoose.astron.nl

\end{tablenotes} 
\end{threeparttable}
\label{table:lsps}
\end{table}
}

\subsubsection{Incoherent Beamforming}
\label{incoherent-beam}

We will compute and process an incoherent sum alongside all our observations in addition to forming coherent beams on individual stars. In the case of an incoherent sum the entire primary field of view, approximately one square degree in L-band, is observed. The computational cost is also much smaller than tiling the primary field of view with coherent beams. However, a disadvantage is that the sensitivity is reduced by a factor of the square root of the number of antennas.

Signals detected in the incoherent sum will be localised after detection, since the raw data are already stored in the voltage buffers. Initially, we will use conventional imaging to localise such signals. It will then be possible to beamform on the sources of the signals at specific locations in the field of view, increasing sensitivity by a factor of $\sqrt{N_{\textrm{ant}}}$, where $N_{\textrm{ant}}$ is the number of antennas in the array. 

\subsubsection{Coherent Beamforming}
\label{beamforming}

BL's commensal survey on MeerKAT is foremost designed to attain the goal of observing one million stars near the Earth. During commensal observations, the nearest stars drawn from our catalog that fall within the primary field of view will be given first priority (with the exception of ad-hoc targets of high importance - see Section~\ref{target_selection}). Each of these stars will be observed individually by means of beamforming for at least five minutes, amounting to a full memory buffer's worth of data at the expected data rate. The duration of five minutes per pointing is consistent with prior BL SETI observations \citep{Price2020, Perez2020, Enriquez2017}, although these prior single-dish surveys adhere to an on-off strategy totaling 15 minutes (3 pointings) on target. Unlike those prior surveys, the on-off strategy is not needed due to advantages of radio telescope arrays and their ability to reject radio frequency interference. For example, widely-separated antennas will tend to downweight nearby sources of RFI when beamforming. We discuss other scenarios for beamforming on stars in Section~\ref{other_scenarios}.  Once all of these stars in a primary field of view have been observed, new observing strategies will be needed.

\begin{figure}
    \centering
    \includegraphics[width=0.7\textwidth]{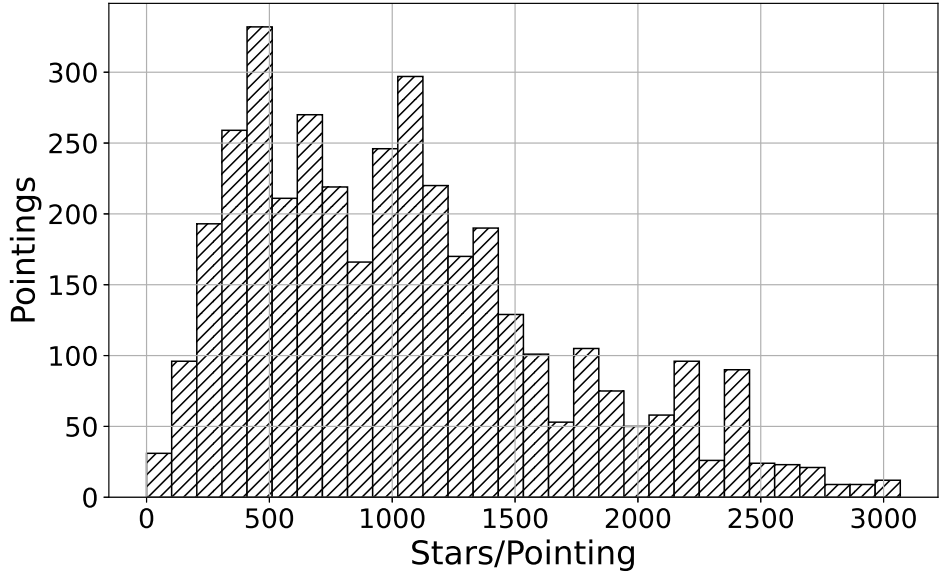}
    \caption{The number of stars within the MeerKAT primary field of view per pointing, for stars drawn from the full catalog. These values are taken from the pointings used in the observing progress simulation discussed in Section~\ref{optimal_timeline}. These pointings are predicted based on information published by the LSPs (see Appendix~\ref{appendix} for full details). In general, there are many more stars per individual pointing than simultaneous beams can be formed: in $\sim 0.8 \%$ of the pointings do the corresponding fields of view contain $\le 64$ stars as drawn from the full catalog. However, many of these pointings are significantly longer than projected processing time (see Table~\ref{table:lsps}), allowing multiple opportunities to observe stars around the same pointing. In addition, typically there is extra time available between LSP observations while the telescope is reconfigured and calibration operations are undertaken. During this time, the buffers still contain the raw voltages from the last pointing. Thus extra opportunities are available for beamforming on stars that were present within the field of view during the last pointing. For further details see Appendix~\ref{appendix}}. Overlapping and repeated observations of the same pointing coordinates are included in this plot. 
    \label{fig:beams_by_LSP}
\end{figure}

The number of stars that can be observed individually via beamforming is limited by several factors. The density of the field itself, in terms of stars drawn from the catalog, dictates the upper limit on the number of stars that may be observed. There is a computational limit to the number of beams that can be formed and processed concurrently while keeping up with the incoming data from MeerKAT. We will be able to form and process 64 concurrent beams at the maximum F-engine data rate for the L-band receivers. Technical details will be described by MacMahon et al.\ (in prep). Many fields of view on the sky will contain far more stars than can be observed with 64 beams. Fig.~\ref{fig:beams_by_LSP} provides an estimate of the distribution of the stellar density in the known LSP fields of view. 
Even though some fields of view will contain far more stars than we can form beams, it may still be possible to observe all of them in practice. Many individual primary telescope pointings will last significantly longer than 5 minutes (see Table~\ref{table:lsps}). Therefore, there will often be a significant amount of time available in which to beamform on all the stars in the field of view while the telescope remains observing the same pointing. For a particular pointing, the expected number of observed stars within the primary field of view $N_{\textrm{obs}}$ is  $\tau_{\textrm{pointing}}N_{\textrm{beams}}/(\tau_{\textrm{rec}} + \tau_{\textrm{proc}})$ for a pointing duration $\tau_{\textrm{pointing}}$, a desired observation time per star of $\tau_{\textrm{rec}}$, processing time of $\tau_{\textrm{proc}}$ and for $N_{\textrm{beams}}$ available beams.

In some cases, it will not be possible to observe all the targets present in each field in this manner. For example, pulsar timing observations may consist of a series of short-duration pointings in sequence. Initially, we will process as many stars as time allows for each pointing, as described in Fig.~\ref{fig:scheduling}. It is possible that the next pointing might offer sources that are less desirable than those within the current field of view which remain unobserved. Therefore, in future, code will be written to automatically determine whether to skip the remaining sources within the current field of view and move on to the next pointing, or to ignore the next pointing and complete processing of the remaining sources within the current field of view. This code will also be able to optimise for a particular figure of merit, such as the one described in Section~\ref{figures_of_merit}. For example, if the current field of view encompasses sufficient unobserved stars which are nearer than those offered by the subsequent pointing, the former will be prioritised at the expense of the latter.

We anticipate that our sample will encapsulate most high priority targets according to a variety of metrics such as planet population or their habitability. Choosing to adopt a ranking based simply on distance and having an expansive catalog ensures most such targets will be properly included. If a given field of view encompasses more stars than can be observed, we will observe as many as possible in order of distance. Another consideration is the projected number of opportunities to observe a particular target, given prior knowledge of upcoming observations. If such information is made available in advance, we will prioritise stars with the fewest expected observing opportunities. 
    
The previous paragraphs have dealt with the problem of fields of view encompassing too many stars for individual beamforming. However, the opposite problem, that there will be more time available for the required observations than there are stars in the field of view, is also likely to occur.  LADUMA \citep{bly2016}, MHONGOOSE \citep{deblock2016} and the MeerKAT Fornax Survey \citep{ser2017} are examples of LSPs which include long-duration observations of small regions of the sky. Despite the number of stars likely to be available within the field of view (see Fig.~\ref{fig:beams_by_LSP}), these observations are so long that these stars will all have been exhausted early on. Alternative strategies will be needed for the remaining observing time in these fields. Our current approach is simply to observe the same stars in the original catalog for longer periods of time. An advantage of this approach is the extension of the ``haystack" \citep{wrightj2018} fraction in time, placing better constraints on the duty-cycle of any potential transmissions. 

The single field of view observed by the LADUMA LSP \citep{bly2016} will allow for possibly the most sensitive SETI experiment ever conducted due to the long observation duration. It therefore presents a unique opportunity for a deep SETI survey. As the pointing is outside of the galactic plane, the field of view is not as densely populated with stars as many of the others. Six stars from the nearest million stars and 183 stars from the full catalog fall within the field of view (when estimated for $\lambda = 21$\,cm). If one were to beamform continuously on the nearest 64 stars within the field of view, the minimum detectable effective isotropic radiated power (EIRP) would range from $1.40 \times 10^{11}$\,W for the nearest of these stars, to $5.57 \times 10^{12}$\,W for the most distant (assuming a signal to noise ratio of 10 and a channel bandwidth of 1\,Hz). Another option would be to tile a region of the field of view with coherent beams. However, 64 beams would cover only a very small fraction of the field of view. Even removing the longer baselines by selecting only the array core, 64 beams would still cover $\ll 1\%$ of the full field of view \citep{hou2021}. Therefore, we do not plan to use such an approach.

\subsubsection{Other Primary Observing Scenarios}
\label{other_scenarios}

\noindent
\textbf{Multiple simultaneous subarrays ---} MeerKAT is capable of running several different observations simultaneously by dividing the full array into smaller independent subarrays. For subarrays, the observing time per star, $\tau_{\textrm{star}}$, will be lengthened as needed to meet equivalent sensitivity levels in order to maintain consistent EIRP constraints on any potential transmissions from stars of equivalent distance. It should also be noted that although there is a $\tau_{\textrm{star}}$ penalty for smaller subarrays, an increased number of beams can be formed, as the number of streams from the F-engines is reduced. The theoretically achievable signal-to-noise ratio is proportional to the square root of the number of antennas and the square root of the observation duration. As given by the radiometer equation:

\begin{equation}
\centering
   \begin{split}
   \frac{S}{N} = \frac{T_{\textrm{src}}}{T_{\textrm{sys}}}\sqrt{\tau_{\textrm{star}}\Delta \nu N_{\textrm{ant}}}
   \end{split}
\end{equation}

\noindent
$\frac{S}{N}$ is the signal to noise ratio, $T_{\textrm{sys}}$ is the system temperature, $T_{\textrm{src}}$ is the source temperature, $\Delta\nu$ is the bandwidth, $\tau_{\textrm{star}}$ is the observation duration and $N_{\textrm{ant}}$ is the number of antennas. To maintain a signal to noise ratio equivalent to that achieved with, for example, 58 antennas (our notional threshold for a full array) and $\tau_{\textrm{star}} = 300$\,s, the observation duration must be increased as follows:

\begin{equation}
   \begin{split}
    \tau_{\textrm{star}}^{\prime} & = \frac{\tau_{\textrm{star}} N_{\textrm{ant}} }{N^{\prime}_{\textrm{ant}}} \\
                          & = \frac{300 \cdot 58}{N^{\prime}_{\textrm{ant}}}
     \end{split}
\end{equation}

\noindent
The new observation duration for $N^{\prime}_{\textrm{ant}}$ antennas is given by $\tau_{\textrm{star}}^{\prime}$. For example, if this 58-antenna array is split into two independent subarrays of 29 antennas each, there would need to be a $2\times$ increase in $\tau_{\textrm{star}}$. 

\noindent
\textbf{Wind stow ---} Given high winds, MeerKAT's antennas enter a wind-stow position, in which their azimuth and elevation remain fixed. It may in future be possible to record data under these conditions, conducting drift-scan observations.   

\subsection{Optimal Timeline}
\label{optimal_timeline}

To evaluate progress towards our stated goals of observing one million nearby stars by means of coherent beamforming, we simulate observing progress based on published observing strategies for the primary observers on MeerKAT. Reasonable estimates for the sky positions of the LSPs can be made using information in available publications (these are provided in Table~\ref{table:lsps}). Table~\ref{table:lsps} also provides an overview of the different LSPs. We use this information to predict our observing progress measured in the number of stars observed over time (Figure~\ref{fig:progress_prediction}). The predictions use known observing strategies, specific pointing positions, and their durations. For a detailed discussion on the procedure used to generate this plot, see Appendix~\ref{appendix}. It should be noted that several LSPs have already begun observing, so some of the pointings used to generate this plot may have already been observed. However, the overall rate of observing progress is not likely to differ greatly, since the telescope time will still be occupied with the remaining observations for each LSP. By drawing from our full sample of stars distributed across the whole sky visible to MeerKAT, we will be able to adapt to new observing programs which have not yet been defined. 

\begin{figure}
    \centering
    \includegraphics[width=0.7\textwidth]{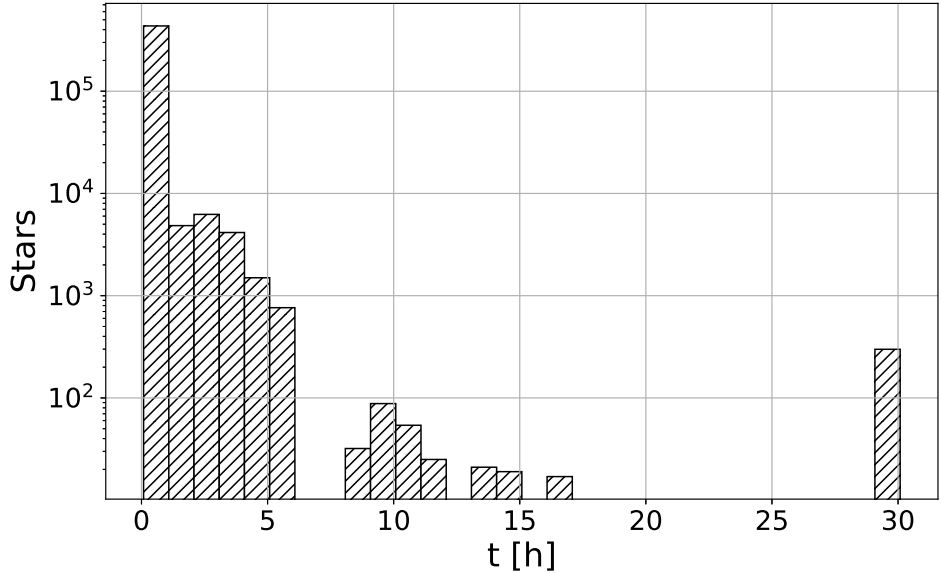}
    \caption{The distribution of observation duration for individual stars after 6 months of simulated observing. The stars with the longest observation durations are from the field of view surrounding the LADUMA pointing.}
    \label{fig:obs_distances}
\end{figure}

\begin{figure}
    \centering
    \includegraphics[width=0.7\textwidth]{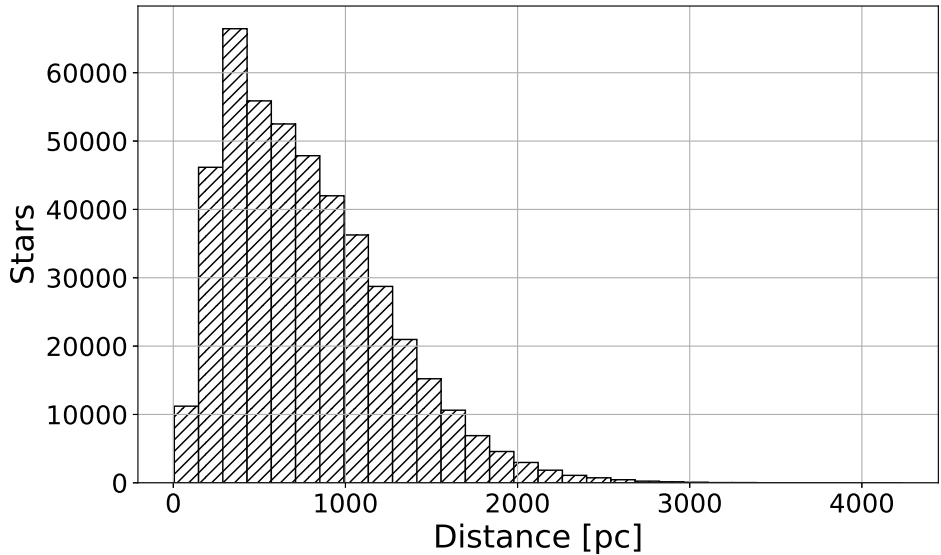}
    \caption{The distribution of distances to the individual stars observed after 6 months of simulated observing.}
    \label{fig:obs_durations}
\end{figure}

\begin{figure}
    \centering
    \includegraphics[width=0.7\textwidth]{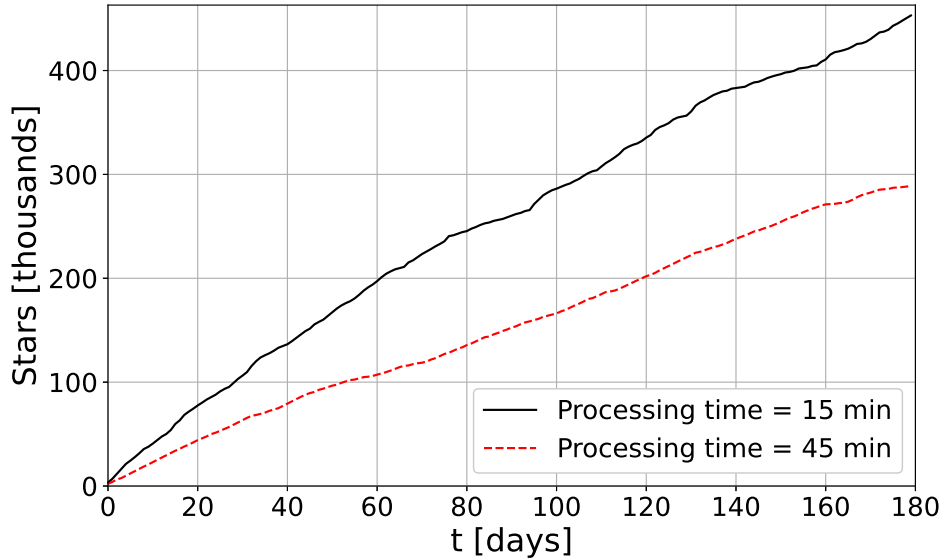}
    \caption{Predicted cumulative observing progress over time given available and estimated pointing information for primary observations. For full details on how progress is estimated, please see Appendix~\ref{appendix}. 
    Each curve illustrates the cumulative number of stars observed over time for processing durations of 15 and 45 minutes. Two banks of processing nodes will be operating in parallel, each of which separately requires the processing time quoted (see Fig.~\ref{fig:scheduling}). Approximately 453000 unique stars are estimated to have been observed after 6 months of commensal observing, for an average of about 2517 per day.}
    \label{fig:progress_prediction}
\end{figure}

Given that $\sim84\%$ (843980) of the nearest one million stars are distributed over the entire sky as visible to MeerKAT, observing progress is mostly dependent on the area of sky covered by the primary LSP observers (Figure~\ref{fig:lsp_pointings}). In the first 6 months of simulated observing, approximately 20900 of the nearest million stars visible to MeerKAT are observed. In order to increase this number, a future primary observing program with large sky-coverage (such as an all-sky survey) would be highly advantageous.  Considering our full catalog, 453000 unique stars may be observed over 6 months, assuming a conservative estimate of 15 minutes of processing time per 5 minutes of buffered raw voltages. Figure~\ref{fig:obs_durations} provides a histogram illustrating the durations for which these stars are estimated to have been observed after 6 months of observing, and Figure~\ref{fig:obs_distances} illustrates the distances to these stars. If a primary observation is shorter than 5 minutes, stars available within the primary field of view will still be observed - but these observations will not be considered ``completed". A full uninterrupted 5-minute observation of each star is required.

\subsection{Evaluating Progress}
\label{obs_efficiency}

We will evaluate our own commensal observing progress by considering the fraction of on-source observing conducted by primary telescope users, for which we recorded and processed data. In making this calculation we do not consider slew time, calibration, maintenance and other factors as on-source observing time since we have no control over them. Rather, we consider on-source observing time as time spent (by a primary observer) on receiving useful data (for example, tracking a particular source). In the case of coherent beamforming, this metric does not take into account missed targets in fields of view for which $N_{\textrm{star}}$ (the number of stars available for observation in a given field of view) $\ge N_{\textrm{obs}}$ (the number of stars that can be observed and the resultant data processed for a given pointing duration). 

\subsection{Target Selection During Observations}
\label{target_selection} 

\begin{figure}
    \centering
    \includegraphics[width=0.7\textwidth]{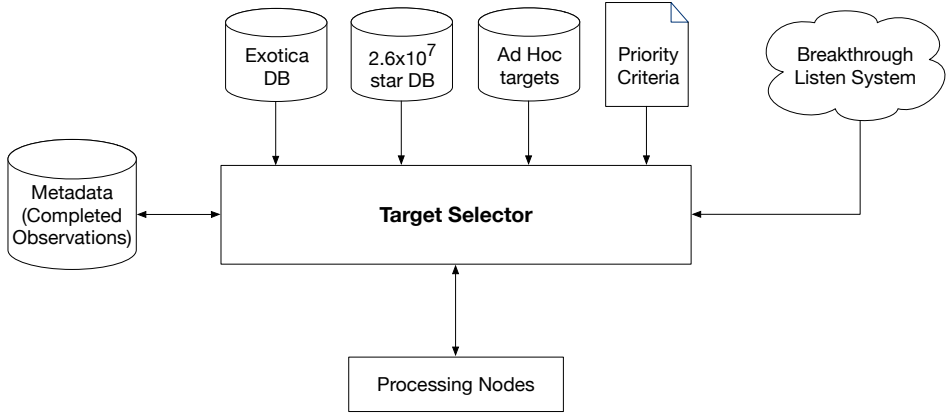}
    \caption{A block diagram of the commensal target selector. Information on the current observation is delivered continuously from the rest of the BL backend to the target selector process. New targets are drawn from the different databases and, based on the current priority criteria, selected for beamforming. The details of these targets are published to the processing nodes.}
    \label{fig:target_selector}
\end{figure}

  An important component of BL's commensal data processing unit on MeerKAT is the target selector.  The target selector process reacts to the current, immediate state of the telescope, providing a ranked list of objects for observation within the field of view. Target selection and ranking takes of the order of seconds for a particular field of view.  This section is concerned only with the mechanism for selecting the targets rather than the beamforming calculations and implementation, which will be described in detail in an upcoming instrumentation paper. Fig.~\ref{fig:target_selector} illustrates the components of the target selector, which is implemented in Python and run as a background process along with the rest of the BL backend. The target selector process interfaces with the target list databases, a database for metadata pertaining to completed observations, and the rest of the BL backend. Via the BL backend, detailed information about the telescope's current state is accessible to the target selector. Targets for beamforming within the current field of view are drawn from the full star catalog. By current field of view, we mean an area of sky encompassed by the full width half maximum (FWHM) or half-power point of the primary beam. A second dynamic catalog will be maintained for ad-hoc targets of high importance (see Section \ref{special_targets}) for immediate observation, should the opportunity arise during commensal observations. These ad-hoc targets will be prioritised over those drawn from all the other catalogs.   

  The immediate observing strategy for the current field of view during observations is determined by a set of priority criteria in conjunction with the database of metadata associated with completed observations. Broadly, observing priority is as follows:

  \begin{enumerate}
    \item Ad-hoc sources of high importance. 
    \item Unobserved sources from the full sample of 26 million stars (ordered by distance).
    \item Unobserved sources from other supplemental catalogs (e.g. exotica).
    \item Sources from the full sample which have already been observed, but for $<5$ minutes or with a subarray containing $<58$ antennas. 
    \item Sources from the full sample that have already been observed, but in a different band. 
    \item Previously observed sources, ordered by distance. 
\end{enumerate}

  Sources such as pulsars and masers will be observed as standard practice and we will use these observations for calibration and to verify the performance of beamforming and other pipeline components. A maximum of one observing slot per pointing (equivalent to a single beam for 5 minutes) will be dedicated to the observation of such sources. The highest priority is given to ad-hoc sources of high importance. These are sources which have been added manually and could, for example, include stars from which particularly interesting signals have been observed in the past. Once the set of targets for beamforming has been determined, they are delivered (ranked by observing priority) to the processing nodes. 

\subsection{Comparison with prior surveys}
\label{figures_of_merit}

We compare the survey planned in this work with recent and historical SETI surveys. Ultimately, MeerKAT will offer at least UHF, L and S-band receivers, but for the purposes of this calculation, we consider only L-band. The continuous waveform transmitter rate figure of merit (TFM), introduced by \cite{Enriquez2017}, is calculated as follows:

\begin{equation}
    \textrm{TFM} = \eta \frac{\textrm{EIRP}_{\textrm{min}}}{\textrm{N}_{\textrm{stars}} \nu_ {\textrm{rel}}}
\end{equation}

\begin{figure}
    \centering
    \includegraphics[width=0.7\textwidth]{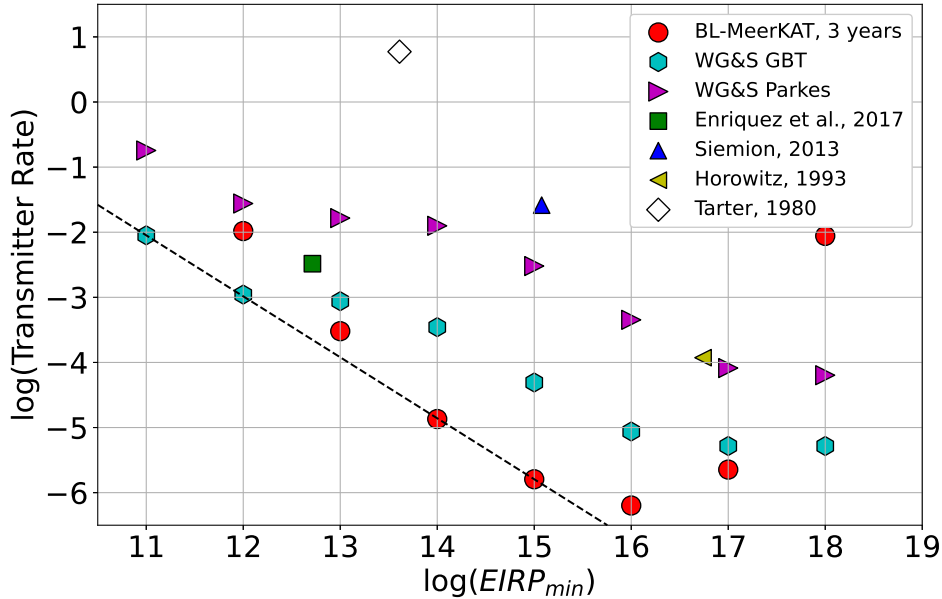}
    \caption{The continuous waveform transmitter figure of merit for stars grouped by their minimum EIRP levels, alongside values for several historical surveys. Some of the historical surveys are represented by a single marker. The values for the GBT and Parkes (labelled WG\&S GBT and WG\&S Parkes) are taken from \cite{Wlodarczyk-Sroka2020}, while the values for BL-MeerKAT are calculated by extrapolating from the observing simulation detailed in Section~\ref{optimal_timeline}. The values presented by \cite{Wlodarczyk-Sroka2020} use a Gaussian distribution to approximate the response of the beams for Parkes and the GBT. As we have limited information on the precise pointings that MeerKAT will undertake during primary science observations, we have assigned each star a randomly selected distance $d$ from the center of the beam, where $0 \le d \le \theta_{b}$ and $\theta_{b}$ is the FWHM at the center frequency $\nu_{c}$. We have used a cosine-squared power estimate for the response of the beam, as recommended in the MeerKAT online documentation (https://science.ska.ac.za/meerkat). The diagonal dotted line indicates the most constraining points in the diagram (after \cite{Enriquez2017}).}
    \label{fig:cwtfm}
\end{figure}

$\textrm{EIRP}_{\textrm{min}}$ is the minimum equivalent isotropic radiated power that can be detected and $\textrm{N}_{\textrm{stars}}$ is the number of stars observed (for the same $\textrm{EIRP}_{\textrm{min}}$ or better). $\eta$ is a normalisation factor, and $\nu_{\textrm{rel}}$ is the bandwidth $\Delta \nu_{\textrm{tot}}$ over the center frequency $\nu_{\textrm{c}}$. The ``transmitter rate" is given by $\frac{1}{\textrm{N}_{\textrm{stars}} \nu_ {\textrm{rel}}}$. \cite{Enriquez2017} used a normalisation factor such that the TFM = 1 when $\nu_ {\textrm{rel}} = 0.5$, $\textrm{N}_{\textrm{stars}} = 1000$ and $\textrm{EIRP}_{\textrm{min}}$ is the EIRP of the former Arecibo Observatory planetary radar at $10^{13}\,\textrm{W}$. In Fig.~\ref{fig:cwtfm} a normalisation factor is not applied when plotting $\textrm{log(transmitter rate)}$ vs $\textrm{log(}\textrm{EIRP}_{\textrm{min}}\textrm{)}$ in keeping with the approach followed by \cite{Enriquez2017} and \cite{Wlodarczyk-Sroka2020}.
The $\textrm{EIRP}_{\textrm{min}}$ for a particular star is proportional to its distance squared: 

\begin{equation}
\textrm{EIRP}_{\textrm{min}} = 4\pi d^2 
F_{\textrm{min}}. 
\end{equation}

$F_{\textrm{min}}$, in W/m$^2$, is given as follows:

\begin{equation}
F_{\textrm{min}} = \frac{S}{N_{\textrm{min}}}\frac{2k_{B}T_{\textrm{sys}}}{A_{\textrm{eff}}}\sqrt{\frac{B}{n_{\textrm{pol}}\tau_{\textrm{obs}}}}
\end{equation}

\noindent
where the desired signal-to-noise threshold is given by $\frac{S}{N_{\textrm{min}}}$, $k_B$ is the Boltzmann constant,  $T_{\textrm{sys}}$ is the system temperature, $A_{\textrm{eff}}$ is the effective collecting area, $B$ is the bandwidth of each channel, $n_{\textrm{pol}}$ is the number of polarisations and $\tau_{\textrm{obs}}$ is the observation length. 

In calculating this result, we group stars by their minimum EIRP values, in a similar manner to the approach described by \cite{Wlodarczyk-Sroka2020}. For each group, the farthest star's $\textrm{EIRP}_{\textrm{min}}$ value is taken.  When calculating each star's $\textrm{EIRP}_{\textrm{min}}$, we consider both distance and total predicted observing time. While \cite{Wlodarczyk-Sroka2020} use a Gaussian distribution to estimate the response of the beam for GBT and Parkes, for MeerKAT, we use a cosine-squared function as recommended in the online MeerKAT documentation\footnote{https://science.ska.ac.za/meerkat}.
We use a desired $\frac{S}{N_{\textrm{min}}}$ of 10, consistent with the work of \cite{Price2020} and \cite{Wlodarczyk-Sroka2020}. Figure~\ref{fig:cwtfm} compares the results obtained by \cite{Wlodarczyk-Sroka2020} for prior BL surveys with the GBT and Parkes. Given the conservative estimate of observing progress over 3 years of operation, it is clear from Fig.~\ref{fig:cwtfm} that the BL-MeerKAT survey will provide a rapid and unprecedented expansion in the number of stars examined for evidence of extraterrestrial intelligence, in many cases obtaining the best $\textrm{EIRP}_{\textrm{min}}$ constraints yet achieved. As a commensal survey, the search itself will also be resource efficient and cost-effective. 

\section{Summary}
\label{conclusion}

 Drawing from the Gaia DR2 database, we have compiled a catalog of approximately 26 million stars for use in BL's commensal SETI survey with MeerKAT. By using parallax and magnitude quality criteria in our distance measurements, we ensure our stellar sample populates a well understood parameter space in the color-magnitude diagram. Our primary stellar sample, reaching up to 3000\,pc, consists mostly of solar type stars with small fractions of cool stars ($T_{\textrm{eff}} < 3700$\,K) and hot stars ($T_{\textrm{eff}} > 6000$\,K). During commensal observations alongside Large Survey Projects, we anticipate that those which observe large solid angles on the sky will contribute more to our pursuit of observing  one million stars. Projects focused on small parts of the sky for many hundreds of hours will be used to undertake some of the most sensitive SETI measurements ever taken. 
 
 With our observing hardware and data collection system directly integrated with the MeerKAT systems, we will make immediate decisions on target selection to realize our priorities. Real time data acquisition on BL hardware will lead to the comprehensive analyses of at least 64 concurrent coherent beams on the sky. For each pointing, we will coherently beamform on as many stars as processing time allows. Alongside coherent beamforming, we will also operate an incoherent mode to take advantage of the full primary field of view of MeerKAT. 

 We anticipate that it will take approximately 1 year and 1 month of commensal observing with MeerKAT to attain our goal of having analysed one million nearby stars. Our observing progress simulations predict that the BL survey on MeerKAT will significantly improve on prior surveys, with the potential to become the most comprehensive SETI survey of its type yet conducted. The search for technologically generated signals, especially those which are narrowband with non-zero drift rates indicative of an extraterrestrial origin, will proceed with unprecedented speed, efficiency and sensitivity.

\acknowledgements{Acknowledgments:  Breakthrough Listen is managed by the Breakthrough Initiatives, sponsored by the Breakthrough Prize Foundation. This work made use of the SIMBAD database (operated at CDS, Strasbourg, France) and NASA's Astrophysics Data System Bibliographic Services. Funding for \BL research is sponsored by the Breakthrough Prize Foundation. 

This work has made use of data from the European Space Agency (ESA) mission Gaia (\href{https://www.cosmos.esa.int/gaia}{link}), processed by the Gaia Data Processing and Analysis Consortium (DPAC, \href{https://www.cosmos.esa.int/web/gaia/dpac/consortium}{https://www.cosmos.esa.int/web/gaia/dpac/consortium}). Funding for the DPAC has been provided by national institutions, in particular the institutions participating in the Gaia Multilateral Agreement.}

\begin{appendices}

\section{Progress Estimation}
\label{appendix}

This appendix explains how a rough estimate of future observing progress is calculated. This estimation is used to produce the plots in Figs.~\ref{fig:beams_by_LSP}, \ref{fig:obs_distances}, \ref{fig:obs_durations} and \ref{fig:progress_prediction}. Observing progress is predicted by considering a simplified, representative day of observing, using information on pointings planned by the main operating primary observers. Pointing information is not always available in advance, such as in the case of Open Time and Director's Discretionary Time, which together account for 33\% of the time on the telescope. The objective of this analysis is to provide a general estimate of observing progress over time, rather than to analyse the precise schedule of observations planned for MeerKAT. Such a schedule is in any case not available months in advance. The representative day of observing is apportioned as follows:

\begin{itemize}
    \item On-sky, science observing time (not including maintenance and other engineering activities) is estimated to amount to 60\% of the day.
    \item The time between LSP observing sessions (for e.g. array reconfiguration and calibration) is estimated to be 20 minutes.
    \item The remaining time is apportioned as follows:
    \begin{itemize}
        \item Approx. 10\% ($\sim 80$ minutes) is allocated to a repeated single pointing (e.g. LADUMA).
        \item Approx. 65\% ($\sim 535$ minutes) is allocated to a long duration single pointing (such as those undertaken by MHONGOOSE and Fornax). On rare occasions a few of these pointings may be repeated (e.g. in the case of MHONGOOSE). 
        \item Approx. 25\% ($\sim 200$ minutes) is allocated to short duration (10 minute) pointings as might be the case for pulsar timing observations or survey projects. 
    \end{itemize}
\end{itemize}

When running the simulation, a star is considered observed if a beam can be formed on it for a full 300\,s (approximately the size of the raw voltage buffers at full datarate). Observing a star in any of the operational observing bands is counted as a completed observation. When the time following an observation is to be used for non-science operations such as array reconfiguration or maintenance, it is allocated for additional processing of the most recent pointing. This is possible because the buffers will still contain the raw voltages recorded during the last pointing. Two banks of processing nodes are assumed to be operating simultaneously in parallel, as described in Fig.~\ref{fig:scheduling}. For each bank of processing nodes, we calculate observing progress for two estimates of processing time ($\tau_{\textrm{proc}}$ = 15 minutes as a conservative estimate and 45 minutes as a worst-case scenario. The daily progress estimate for each bank is added together to produce Fig.~\ref{fig:progress_prediction}. The pseudocode for calculating observing progress per day is given below: 

\begin{algorithmic}
\FOR {each day}
    \FOR {each LSP observing session}
         \FOR {each pointing of duration $\tau_{\textrm{point}}$}
             \STATE { $N_{\textrm{rec}} = \tau_{\textrm{point}}//(\tau_{\textrm{proc}} + \tau_{\textrm{rec}})\cdot N_{\textrm{beams}}$}
             \IF {$N_{\textrm{star}} \le N_{\textrm{rec}} $}
                 \STATE { $N_{\textrm{day}} = N_{\textrm{day}} + N_{\textrm{star}}$ }
            \ELSE
                \STATE { $N_{\textrm{proc}} = (\tau_{\textrm{point}}\%(\tau_{\textrm{proc}} + \tau_{\textrm{rec}}) + \tau_{\textrm{post}})//\tau_{\textrm{proc}}\cdot N_{\textrm{beams}} $ }
                \STATE { $N_{\textrm{post}} = \textrm{MIN}(N_{\textrm{proc}}, N_{\textrm{star}} - N_{\textrm{rec}})$ }
               \STATE { $N_{\textrm{day}} = N_{\textrm{day}} + N_{\textrm{rec}}$ }
               \STATE { $N_{\textrm{day}} = N_{\textrm{day}} + N_{\textrm{post}}$ }
            \ENDIF
         \ENDFOR
     \ENDFOR
 \ENDFOR
 \end{algorithmic}
 \hfill \break

\noindent
$N_{\textrm{star}}$ is the number of as-yet unobserved stars in the current field of view, and $N_{\textrm{day}}$ is the number of new, unique stars observed each day. $N_{\textrm{beams}}$ is the number of beams that can be formed simultaneously. $N_{\textrm{rec}}$ is the number of slots available for both recording and processing for a pointing of duration $\tau_{\textrm{point}}$. $\tau_{\textrm{rec}}$ is the recording time, and $\tau_{\textrm{proc}}$ is the processing time for $\tau_{\textrm{rec}}$'s worth of recorded data. $\tau_{\textrm{post}}$ is the length of time between the end of the current pointing and the beginning of the next (for example, the time between LSP observing sessions). $N_{\textrm{proc}}$ is the number of processing slots available in this time and any leftover time after $N_{\textrm{rec}}$ beams have been formed and processed. $N_{\textrm{post}}$ is the actual number of new stars observed during the time before the next pointing begins. 

There are a number of important assumptions that have been made in estimating observing progress. Actual daily observing schedules will differ from the one given above. It is designed merely to be representative of the time allocation and number of pointings we might expect per day. It should also be noted that observing plans may differ in practice from those given by each LSP in the references provided in Table~\ref{table:lsps}. In addition, a number of LSPs have already begun their observations in at least some capacity. Therefore, it is likely that some of the pointings considered in this progress estimation will have already been observed. However, this does not matter when estimating progress in terms of the number of stars observed. Even if the pointings are different in practice, they will likely still offer a broadly similar number of new stars within the field of view for observation (see Fig.~\ref{fig:beams_by_LSP}).  

\end{appendices}

\clearpage

\bibliographystyle{aasjournal}
\bibliography{meerkat_targets.bib}

\end{document}